\documentclass[preprint,12pt]{elsarticle}

\usepackage{amssymb}

\usepackage[tbtags]{amsmath}

\begin{document}

\title{Stark-chirped rapid adiabatic passage in the presence of dissipation for quantum computation}

\begin{frontmatter}

\author[sw,cqt]{X. Shi}
\author[cqt]{C. H. Oh\footnote{phyohch@nus.edu.sg}}
\author[sw,cqt,syu]{L. F. Wei\footnote{weilianfu@gmail.com}}
\address[sw]{Quantum Optoelectronics Laboratory, School of Physics
and Technology, Southwest Jiaotong University, Chengdu 610031,
China}
\address[cqt]{Centre for Quantum Technologies and Department
of Physics, National University of Singapore, 3 Science Drive 2,
Singapore 117542}
\address[syu]{State Key Laboratory of Optoelectronic Materials and
Technologies, School of Physics Science and Engineering, Sun Yet-sen
University, Guangzhou 510275, China}

\begin{abstract}
Stark-chirped rapid adiabatic passage (SCRAP) is an important technique used for coherent quantum controls. In this paper we investigate how the practically-existing dissipation of the system influences on the efficiency of the passage, and thus the fidelities of the SCRAP-based quantum gates. With flux-biased Josephson qubits as a specifical example, our results show clearly that the efficiency of the logic gates implemented by SCRAP are robust against the weak dissipation. The influence due to the non-adiabtic transitions between the adiabatic passages is comparatively significantly small. Therefore, the SCRAP-based logic gates should be feasible for the realistic physical systems with noises.

\end{abstract}


\end{frontmatter}

\section{Introduction}
Over recent years, quantum computation has attracted much
attention partly because the discovery of quantum algorithm for
specific problems provides a tremendous
speedup in computation, compared to a classical computer~\cite{Shor,Grover}. A critical prerequisite for building a
quantum computer is to perform the basic single- and two-qubit gates with high fidelity above certain threshold levels~\cite{Jerry,Norbert}.

A typical ingredient in quantum computing is the coherent transfers of the population between the qubit states. Basically, there are two approaches to realize the population transfers between two selected quantum states; one makes use of the Rabi oscillations and the other is based on population
passages~\cite{Fleischhauer,Shore,Mei}. For Rabi oscillations, the transfer efficiency is strongly dependent on the precisely-designed duration of the applied pulse. On the other hand, the logic gates implemented via population passage techniques, such as shortcut to adiabatic passage~\cite{Mei2}, the stimulated Raman adiabatic passage (STIRAP)~\cite{Bergmann} and the Stark-chirped rapid adiabatic passage (SCRAP)~\cite{Wei}, are evolution-time insensitive and thus robust against the imperfections of durations of the applied pulses.
Until now, most of population passage schemes to implement the quantum computation are based on the pure quantum systems, but their practically-existing dissipative effects (e.g., spontaneous emissions, phase relaxations and the outsides from the system) have not be exactly treated. The fidelities of the logic gates for the realistic quantum computing demonstrations are particularly important, therefore, it is necessary to investigate how the practically-existing dissipation influences on the efficiencies of the population passages and consequently the fidelities of the relevant logic gates for quantum computing.

Usually, the dynamics of an open quantum system can be described by two approaches~\cite{Scully}: the master equation for the reduced density matrix and the Heisenberg-Langevin equation by introducing environment noise operators. Additionally, a relatively-simple approach, i.e., dissipative Schr$\rm \ddot{o}$dinger equation with a non-Hermitian Hamiltonian describing the damping, is also utilized. In this approach the environment effects are considered simply by phenomenologically introducing certain non-Hermitian terms in the Hamiltonian of the system. Then, the dynamics of the treated open system can still be described by the usual Schr\"odinger equation. Indeed, this idea has been utilized to investigate the dissipative  effects in the well-known STIRAP for three-state $\Lambda$ atomic systems~\cite{Vitanov2}, wherein the decay rate $\Upsilon$ of the intermediate state $|2\rangle$ is served as the main source of the dissipation during the population transfers from the state $|1\rangle$ to the target state $|3\rangle$. The damping of the transfer efficiency with $\Upsilon$ shows different behaviors, i.e., exponential at small $\Upsilon$ and polynomial at large $\Upsilon$. This feature provides a realistic STIRAP scheme for three-state $\Lambda$ atomic systems in the presence of decay of the intermediate state.
For the present two-state system, the decay of the excited state is the dominant dissipative source which mainly decreases the transfer efficiency of the SCRAP~\cite{Dridi}. In this brief report, we examine how this dissipation influences the fidelity of the SCRAP-based logic gates. For the simplicity, we treat the dissipation related to the excited state of the qubit by adding an imaginary part to the relevant diagonal term of the non-dissipative Hamiltonian. Our proposal is demonstrated specifically with the flux-biased Josephson qubits, but can also be applied to the other driven solid-state qubit systems.

\section{Definition of the model}
Without loss of the generality, we assume that the two-state system is well prepared initially, at time $t_0=-\infty$, in the ground state $|0\rangle$. Our end is to maximize the final population $P_1(\infty)$ of the target state $|1\rangle$ in the presence of the decay of $|1\rangle$.
Originally, without dissipation of the two-state system the desired transfer can be precisely implemented by means of the standard SCRAP~\cite{Rickes,Rangelov,Yatsenko}. However, due to various practically-existing noises, such a transfer should be influenced.
As a simplified model the state decay of a quantum system can be generically
described by adding a loss rate $\Gamma\,(>0)$ to its non-lossy Hamiltonian, as a negative imaginary part to the corresponding
diagonal term. As a consequence, the time evolution of the probability amplitudes for the dissipative driven two-level system can be expressed by the equation~\cite{Vitanov1}
\begin{align}
i\hbar\frac{d}{dt}\left(\begin{array}{c}
C_0\\
C_1
\end{array}\right)=\frac{\hbar}{2}\left(
          \begin{array}{cc}
          0          &\Omega(t)\\
          \Omega(t)  &2\Delta(t)-2i\Gamma
          \end{array}
              \right)
              \left(\begin{array}{c}
C_0\\
C_1
\end{array}\right).\label{eq:1}
\end{align}
Here, $\Omega(t)$ is the Rabi frequency coupling the levels of the two-state system, $\Delta(t)$ is relative to the pulse chirping the excited level and $C_0(t)$ and $C_1(t)$ are the probability amplitudes related to the states $|0\rangle$ and $|1\rangle$, respectively.

To analyze the progress of SCRAP in the presence of state decay, we define the adiabatic states $|+\rangle=\sin\theta(t)|0\rangle+\cos\theta(t)|1\rangle$ and $|-\rangle=\cos\theta(t)|0\rangle-\sin\theta(t)|1\rangle$, which are the instantaneous eigenstates of the Hamiltonian in Eq.~\eqref{eq:1} with $\Gamma=0$. Here, the mixing angle $\theta(t)$ is defined as $\theta(t)=\arctan[\Omega(t)/\Delta(t)]/2$.
In the basis defined by the adiabatic vectors $|+\rangle$ and $|-\rangle$, equation \eqref{eq:1} can be written as
\begin{align}
i\frac{d}{dt}&\left(\begin{array}{c}
a_+\\
a_-
\end{array}\right)=\frac{1}{2}\left(
          \begin{array}{cc}
          \varepsilon_+(t)        &2i\dot{\theta} \\
          -2i\dot{\theta} &\varepsilon_-(t)
          \end{array}
              \right)
                \left(\begin{array}{c}
a_+\\
a_-
\end{array}\right)\nonumber\\
              &+\left(
          \begin{array}{cc}
          -i\Gamma\cos^2\theta         &i\Gamma\sin\theta\cos\theta \\
          i\Gamma\sin\theta\cos\theta &-i\Gamma\sin^2\theta
          \end{array}
              \right)
              \left(\begin{array}{c}
a_+\\
a_-
\end{array}\right),\label{eq:2}
\end{align}
with $\varepsilon_{\pm}(t)$$=$$\Delta(t)\pm\sqrt{\Delta^2(t)+\Omega^2(t)}$.

Obviously, the off-diagonal elements in Eq.~\eqref{eq:2} result in the coupling between two adiabatic states $|+\rangle$ and $|-\rangle$ (i.e., the passage paths for the desired population transfers). For the ideal case without state decay, i.e.,$\Gamma=0$, the desired adiabatic transfer can be implemented by properly designing the applied pulses to satisfy the condition: $\dot{\theta}=0$, i.e., the adiabatic condition~\cite{Shi,Nie}
\begin{align}
\eta=\frac{\left|\Omega(t)d\Delta(t)/dt-\Delta(t)d\Omega(t)/dt
\right|}{2[\Delta^2(t)+\Omega^2(t)]^{3/2}}\ll 1.
\end{align}
However, the second term in Eq.~\eqref{eq:2} shows that the damping of SCRAP corresponds to two ways, one is the decay of the adiabatic passage paths described by the diagonal elements and the other is the transition damping described by the nonzero off-diagonal elements.
For a counterintuitive pulse sequence with initial state $|0\rangle$ (at time $t=-\infty$, $\theta=0$, then $\theta=\pi/2$ at $t=\infty$), the transfer progress (which transfers the population from state $|0\rangle$ to state $|1\rangle$) goes along the adiabatic path $|-\rangle$ with the decay rate $\Gamma\sin^2{\theta}$. Along this adiabatic passage, the final population of state $|1\rangle$ is
\begin{align}
P_1^{ci}\approx\exp{\left(-2\Gamma\int_{-\infty}^{\infty}
\sin^2{\theta(t)}dt\right)}, \label{eq:4}
\end{align}
while if the system is initially prepared at the state $|1\rangle$, then the population is transferred along the adiabatic passage $|+\rangle$ (with the decay rate $\Gamma\cos^2{\theta}$) to the ground state $|0\rangle$. The final population of the state $|0\rangle$ reads
\begin{align}
P_0^{ci}\approx\exp{\left(-2\Gamma\int_{-\infty}^{\infty}
\cos^2{\theta(t)}dt\right)}.\label{eq:5}
\end{align}
Absolutely, the non-adiabatic transition between the states $|-\rangle$ and $|+\rangle$ may also lead to the losses of $P_1^{ci}$ and $P_0^{ci}$. As the dissipation is irreversible, the population transfer may be significantly destroyed by the strong dissipation.

\section{quantum logic gates in the presence of dissipation}
In what follows, we investigate specifically how the dissipation of system influences the fidelity of the SCRAP-based quantum logic gates. Our discussion is based on the SCRAPs in flux-biased Josephson qubits, but can be easily generalized to other physical systems. For operational simplicity, here linear Stark pulses, rather than the previous gaussian Stark pulses~\cite{Nie}, are applied to the qubits.

The quantum behavior of a flux-biased Josephson junction has been described in detail elsewhere~\cite{Clarke,MartinisB}. The Hamiltonian of the system is
\begin{align}
\hat{H}_s(t)=&\frac{p^2}{2m}+E_J\left(\frac{(\delta-\phi_{b0})^2}{2\lambda}
-\cos{\delta}\right)\nonumber\\
&-\frac{\Phi_0}{2\pi}\left(\frac{M}{L}I_{dc}+I_{ac}\right)\delta.\label{ll}
\end{align}
Here, the pump pulse $I_{ac}=\xi(t)\cos(\omega_{10}t)$ is used to couple the qubit states and the Stark pulse $I_{dc}$ is applied to chirp the qubit's transition frequency $\omega_{10}$. Also, $\Phi_0=h/2e$ is the flux quantum, $E_J=I_0\Phi_0/2\pi$ is the Josephson energy, and $\lambda=2\pi I_0L/\Phi_0$, $m=C_J[\Phi_0/(2\pi)]^2$, $\phi_{b0}=2\pi I_{\phi0}M/\Phi_0$. Consequently, the Hamiltonian of the driven qubit (with decay rate $\Gamma$) in the interaction picture can be expressed as
\begin{align}
\hat{H}_{\rm int}(t)=\left(
          \begin{array}{cc}
          0          &-\frac{\Phi_0}{2\pi}\kappa\delta_{01}\\
          -\frac{\Phi_0}{2\pi}\kappa\delta_{10}  &-\frac{\Phi_0}{2\pi}\Delta_1(t)-i\hbar\Gamma
          \end{array}
              \right).\label{eq:7}
\end{align}
where $\delta_{ij}=\langle i|\delta|j\rangle,\,i,j=0,1$, $\kappa=\xi(t)/2$ and $\Delta_1(t)=MI_{dc}(t)(\delta_{11}-\delta_{00})/L$.
\begin{figure}[htbp]
\includegraphics[scale=0.4]{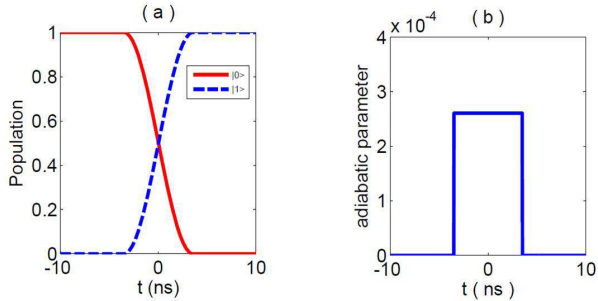}
\centering
\caption{(Color online)The population transfer without dissipation for implementing the single-qubit gate with a flux-biased Josephson junction. In (a) the two pulses are designed with a linear forms $I_{dc}(t)=0.1t~\rm
A$ and $\xi(t)=-1.88~\rm nA~(-3.5~\rm n s\leq t\leq 3.5~\rm n
s,~\text{else where}~\xi(t)=0 ~\rm V/m)$. With this pulse sequence, the system initially prepared in the state
$|0\rangle$ completely transfers to the state $|1\rangle$. The corresponding adiabatic parameter is shown in (b).}
\end{figure}

When $\Gamma=0$, i.e., for the ideal system without dissipation, we show in Fig.~1(a) that the single-qubit gate, i.e., the qubit inversions, can be realized by using a linear pump pulse $I_{ac}$ and a Stark pulse $I_{dc}$ to implement the desirable population transfer between the qubit states. It is shown that, under the counterintuitive pulse sequence (the applied Stark pulse $I_{dc}$ precedes the pump pulse but turns off first), the qubit inversion is realized along the adiabatic passage $|-\rangle$ (with $100\%$ probability). Fig.~1(b) exhibits that the adiabatic parameter $\eta$ is fairly smaller than $1$. This implies that the above progress for population transfers is really confined in the adiabatic region. Unlike the Gaussian pulse used to control the population transfer~\cite{Nie}, the maximum value of the adiabatic parameter reached $120$, thus it is not the adiabatic progress.
Note that the desired population inversions are finished within a relatively-short time interval, i.e., $\tau_1=20\rm ~ns$, which is really rapid compared to the typical decoherence time (e.g., $0.3~\rm \mu s$~\cite{Clarken}).

Now, let us consider how the dissipation of the system influences the above qubit inversions. The decay rate $\Gamma$ is meaningless unless it is related to a real physical variation, such as the characteristic width of the driving pulses $T$. For the convenience, we introduce a dimensionless decay rate $\gamma=\Gamma T$~\cite{Vitanov2} to illustrate the dissipation of our model. Then, the dissipation of the system can be divided into three regions; (i) weak dissipation ($\gamma\ll1$), (ii) strong dissipation ($\gamma\sim 1$) and (iii) very strong dissipation ($\gamma\gg 1$). In Fig.~2 we show how the population probability of the target state varies with
the decay rate $\gamma$ and the evolution time $t$ for the applied counterintuitive sequence pulses. Specifically, Figs.~2(a) and 2(b) illustrate the population passage from the initial state $|0\rangle$ to the target state $|1\rangle$ along the adiabatic passage $|-\rangle$; while Figs.~2(c) and 2(d) are relative to the population transfer from the state $|1\rangle$ to the state $|0\rangle$ along the adiabatic passage $|+\rangle$. The time-dependent population probabilities of the target state are calculated by Eqs.~\eqref{eq:4}, \eqref{eq:5}. As a comparison, we also provide the relevant results by directly solving the Schr$\rm\ddot{o}$dinger equation with Hamiltonian \eqref{eq:7}.
Here, we assume the qubit is in the initial state at time $t_0=-10~\rm ns$, the passage transfer is finished at time $t_f=10~\rm ns$, and the system is in the superposition state during the time $t_{b}=-3.5~\rm ns$ to $t_{m}=3.5~\rm ns$. It is shown that the adiabatic approximation made for delivering Eqs.~\eqref{eq:4} and \eqref{eq:5} works well.
\begin{figure}[htbp]
\includegraphics[scale=0.4]{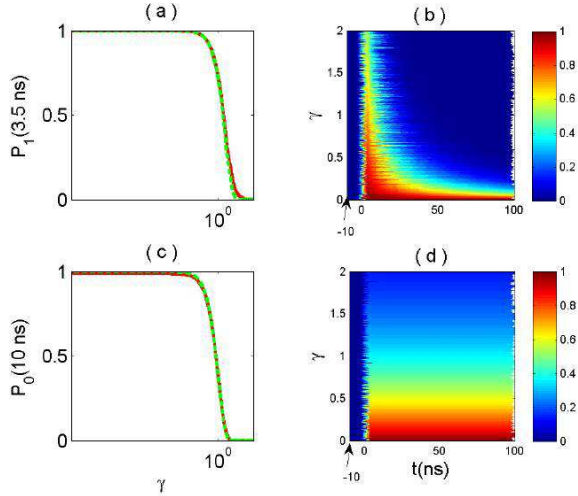}
\centering
\caption{(Color online) Population transfer with different decay rate for single-qubit gate. The pulses used to implement the adiabatic passage are the same linear pulses with the counterintuitive sequence for single-qubit gate discussed without dissipation. The population probability with initial state $|0\rangle$ at time $t_0=\rm -10~ns$ varies with $\gamma=\Gamma T$ ($T=2\times10^{-8} $) described by (a), while (c) is relative to the initial state $|1\rangle$.  The red lines both in (a) and (c) are obtained by numerical solution of the Shr\"odinger equation related to Eq.~\eqref{eq:7}, and the dashed green line in (a) and (c) is the analytical results from Eq.~\eqref{eq:4} and Eq.~\eqref{eq:5} respectively. Finally, (b) and (d) give the dynamics of the population marked with different colors for the varied $\gamma$ and the evolution time $t$ with initial state $|0\rangle$ and $|1\rangle$, respectively.}
\end{figure}
The above numerical results show clearly that:
(i) in the weak dissipation region, i.e., $\gamma\ll 1 $, the dissipation can be really neglected, and the efficiency of the population transfer is sufficiently high (almost $100\%$); in the strong dissipation, i.e., $\gamma\sim 1$, the population efficiency is lower than $1$; the final population may completely destroyed with a very strong dissipation $\gamma \gg 1$.
(ii) For the population passage from the state $|0\rangle$ to the state $|1\rangle$, the transfer probability decreases as an exponential function $\exp[-2\Gamma(t-t_m)]$ after the passage region $t>t_m$; while, for the passage from the state $|1\rangle$ to the state $|0\rangle$, the loss of the population can still be described by an exponential function $\exp[-2\Gamma(t-t_{b})]$ before the passage region $t<t_b$.
(iii)The non-adiabatic transition weakly influences the population transfer, and the dissipation of the system is mainly from the decay of the adiabatic passage paths.

To be more thorough, we investigate how the dissipation influences the SCRAP-based two-qubit gate with two capacitively-coupled flux-biased Josephson qubits. For the simplicity, here the two junctions are assumed to be identical and thus two qubits possess the same energy structure. Originally, the two-qubit gate can be implemented also by the adiabatic population passages~\cite{Wei} via applying a controllable dc current $I^{(2)}_{dc}$ to chirp the second qubit. Considering the practically-existing decay of the excited state of the qubits (with the same decay rate $\Gamma$ for simplicity), the Hamiltonian of such a driven two-qubit system can be simply expressed as
\begin{align}
\hat{H}_{I}(t)=\left(
          \begin{array}{cccc}
          \Delta_{00}          &0    &0     &0\\
          0          &\Delta_{01}-i\hbar\Gamma    &\Omega_{01}     &0\\
          0          &\Omega_{10}    &\Delta_{10}-i\hbar\Gamma     &0\\
          0          &0    &0     &\Delta_{11}-2i\hbar\Gamma
          \end{array}
              \right) \label{eq:8}
\end{align}
with
\begin{align}
\Delta_{00}=&-\frac{M\Phi_0}{2\pi L}I^{(2)}_{dc}(t)\delta_{00}+(\frac{2\pi}{\Phi_0})^2\frac{1}{\bar{C}_m}
p^{(1)}_{00}p^{(2)}_{00},\nonumber \\
\Delta_{01}=&-\frac{M\Phi_0}{2\pi L}I^{(2)}_{dc}(t)\delta_{11}+(\frac{2\pi}{\Phi_0})^2\frac{1}{\bar{C}_m}
p^{(1)}_{00}p^{(2)}_{11},\nonumber\\
\Delta_{10}=&-\frac{M\Phi_0}{2\pi L}I^{(2)}_{dc}(t)\delta_{00}+(\frac{2\pi}{\Phi_0})^2\frac{1}{\bar{C}_m}
p^{(1)}_{11}p^{(2)}_{00},\nonumber\\
\Delta_{11}=&-\frac{M\Phi_0}{2\pi L}I^{(2)}_{dc}(t)\delta_{11}+(\frac{2\pi}{\Phi_0})^2\frac{1}{\bar{C}_m}
p^{(1)}_{11}p^{(2)}_{11}, \nonumber
\end{align}
and
\begin{align}
\Omega_{01}=\Omega_{10}=(\frac{2\pi}{\Phi_0})^2\frac{1}{\bar{C}_m}
p^{(1)}_{10}p^{(2)}_{10} \nonumber
\end{align}
where $\bar{C}_m=C_J(1+\zeta)/\zeta$ ($\zeta$ is the effective coupling coefficient) represents the interaction between two qubits and $p_{ij}=-i\hbar\langle i|\frac{\partial}{\partial \delta}|j\rangle$, $p^{(1)}_{ij}=p^{(2)}_{ij}$.

\begin{figure}[htbp]
\includegraphics[scale=0.6]{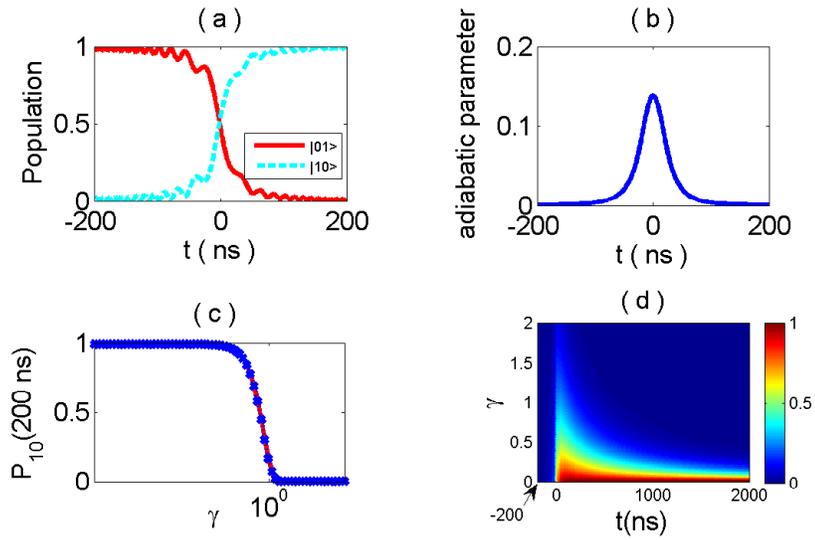}
\centering
\caption{(Color online) Population transfer for the two-qubit gate with a Stark pulse $I^{(2)}_{dc}=-3.5t$. (a) population transfers between the two-qubit states $|01\rangle$ and $|10\rangle$, and (b) the corresponding adiabatic parameter during the passages.
(c) The final population of the state $|10\rangle$ at a defined time $t=\rm 200~ns$ varies with the dissipation $\gamma=\Gamma T$ (with $T=4\times10^{-7} $). The red line in (c) is obtained by numerical solution to the Schr\"odinger equation related to the Hamiltonian~\eqref{eq:8} and the dotted blue line is the analytic solution to the dynamics for the reduced Hamiltonian~\eqref{eq:9}. (d) Probability of transfer from the states $|01\rangle$ to $|10\rangle$ varies with the dissipation parameter $\gamma$ and the evolution time $t$. Colorbar implies the variation of probability.}
\end{figure}

Still, one can easily check that the populations of $|00\rangle$ and $|11\rangle$ of the present two-qubit system are always unchanged, and the population transfer only occurs between the states $|01\rangle$ and $|10\rangle$. So the dynamics of the two qubits can be limited to a $2\times2$ subspace generated by the states $|01\rangle$ and $|10\rangle$. In absence of the dissipation, i.e., $\Gamma=0$, Fig.~3(a) shows that the population transfer can be easily achieved between the states $|01\rangle$ and $|10\rangle$. Fig.~3(b) displays that the maximum value of  the adiabatic parameter $\eta$ during such a passage is about $0.14$. Thus, the usual i-SWAP gate has been realized by the adiabatic SCRAP technique.

In Figs.~3(c) we investigate how the dissipation influences the population transfer from the state $|01\rangle$ to $|10\rangle$ for a defined passage time interval $\tau_2=400~\rm ns$. It is shown that results by numerically solving the Schr$\rm\ddot{o}$dinger equation with the Hamiltonian \eqref{eq:8}, and those by analytically solving the evolution within the subspace with the reduced Hamiltonian (defined by the adiabatic vectors $|+\rangle$ and $|-\rangle$)
\begin{align}
\hat{H}'_{I}(t)=\left(
          \begin{array}{cc}
          \epsilon_+ -i\hbar\Gamma         &0   \\
          0    & \epsilon_- -i\hbar\Gamma
          \end{array}
              \right),\label{eq:9}
\end{align}
with $\epsilon_{\pm}=(\Delta_{10}-\Delta_{01}\pm\sqrt{4\Omega^2_{01}+(\Delta_{10}-\Delta_{01})^2 })/2$ are consistent. Obviously, the dissipation of the two-qubit operation is not relative to the non-adiabatic transition between the two passage paths $|+\rangle$ and $|-\rangle$. Moreover, the dissipation of the SCRAP-based two-qubit gate can be also divided into three regions. The efficiency of the population transfer is sufficiently high in the weak dissipation region $\gamma\ll 1 $, but it is decreasing when the system is in the strong ($\gamma \sim 1$) and very strong dissipation ($\gamma \gg 1$) regions.
In Fig.~3(d) we depict how the transfer probability depends on the dissipation parameter $\gamma$ and the evolution time $t$. We can see from the figure that, for the sufficiently-weak dissipation (typically for $\gamma<0.1$) the passage time could be set as a sufficiently-long interval, e.g., $2\mu s$ (if it is still shorter than the decoherence time of the system). However, for the strong dissipations, $\gamma\sim 1$ and $\gamma \gg 1$, the population transfer should be achieved within sufficiently-short time interval.

\section{Conclusion}
In summary we have investigated the Stark-chirped rapid adiabatic passage (SCRAP) of a driven dissipative two-level system. As a simplified model, we describe the dissipation of the system by adding a phenomenal parameter $\Gamma$ to the chirped excited state of the system.
Then, by solving the relevant Schr\"odinger equation we then discuss how the practically-existing dissipation influences the population transfer between the two selected levels of the system. We have found that the desired SCRAP probability is related to the effective dissipative parameter $\gamma=\Gamma T$ (with $T$ being the time interval of population passage), and consequently we can divided the dissipation into three regions; (i) weak dissipation ($\gamma\ll1$), (ii) strong dissipation ($\gamma\sim 1$) and (iii) very strong dissipation ($\gamma\gg 1$). In the weak dissipation region ($\gamma\ll1$), the interaction between the quantum system and the environment is really small, thus the influence from the environment is sufficiently weak. As  a consequence, the population transfer from the initial state to the target state can be robustly implemented. As the interaction between the quantum system and the environment increases ($\gamma\sim1$), the leakage of the quantum system increases, such that the population probability is decreasing. When the coupling between the quantum system and environment is very strong ($\gamma\gg1$), the situation is more complex: (i) If the qubit is initially prepared at its ground state, the effect of the large decay rate makes the quantum system decouple from the controlling pulses (pump pulse and Stark pulse), then the qubit will not be excited to its excited state and is still in its initial ground state; (ii)If the qubit is initially prepared at the excited state, the relevant population will decay quickly to the environment and the system could not be excited again. Our numerical results clearly show that, in the weak dissipation regime, the SCRAP-based quantum computing scheme still works well; while in the strong dissipation regime the fidelity of quantum gate implemented by the SCRAP technique decreases manifestly. Certainly, if the system works in the very strong dissipation regime, then the SCRAP technique can not be utilized to implement quantum computing.

Our generic discussion has been demonstrated with a typical quantum computing system, i.e., the flux-biased Josephson qubits. In this specific model we have found that the loss of the transfer efficiency of the SCRAP is related to both the non-adiabatic transitions between the adiabatic passage paths and the decay of the adiabatic passage paths. During the passage for implementing the  single-qubit gate, we find that the loss owing to dissipation-induced  transition between two adiabatic passage paths is really small and thus negligible. For the two-qubit gate, we find that the dissipation-induced  transition between two adiabatic passage paths vanish, and only the decay of the adiabatic passage paths exists. Based on this analysis we have delivered a proper approach to implement the quantum logic gates in such a system in the presence of dissipation. Our results provide quantitative estimates of the population losses during the SCRAPs, and thus should be useful for the realistic qubit operations.

{\bf Acknowledgements}: This work was supported in part by the National Science Foundation
grant Nos. 90921010, 11174373, the National Fundamental Research
Program of China through Grant No. 2010CB923104, National Research Foundation and Ministry of Education, Singapore (Grant No. WBS: R-710-000-008-271), the 2013 Doctoral Innovation funds of Southwest Jiaotong University and the Fundamental Research Funds for the Central Universities.

\label{}

\bibliographystyle{elsarticle-harv}
\bibliography{<your-bib-database>}

\end{document}